\documentstyle[preprint,floats,pra,aps,psfig]{revtex}

\newcommand{\doublespace}{
    \renewcommand{\baselinestretch}{1.6}\large\normalsize}

\newcommand{\bce}{\begin{center}}
\newcommand{\ece}{\end{center}}
\newcommand{\be}{\begin{equation}}
\newcommand{\ee}{\end{equation}}
\newcommand{\bea}{\vspace{0.25cm}\begin{eqnarray}}
\newcommand{\eea}{\end{eqnarray}}

\def\PLA{{Phys. Lett.}  A }
\def\PLB{{Phys. Lett.}  B }
\def\PRL{{Phys. Rev. Lett.} }
\def\PRA{{Phys. Rev.} A }

\def\PRD{{Phys. Rev.} D }

\doublespace

\begin{document}

\title{{\LARGE {\bf About entanglement properties of kaons and tests of hidden variables models}}}

\doublespace

\author{M.Genovese \footnote{ genovese@ien.it. Tel. 39 011 3919253, fax 39 011 3919259}}
\address{Istituto Elettrotecnico Nazionale Galileo Ferraris, Str. delle Cacce
91,\\I-10135 Torino. }
\maketitle

\vskip 1cm {\bf Abstract} \vskip 0.5cm In this paper we discuss
entanglement properties of neutral kaons systems and their use for
testing local realism. In particular we analyze a Hardy-type
scheme \cite{Garb} recently suggested for performing a test of
hidden variable theories against standard quantum mechanics. Our
result is that this scheme could in principle led to a conclusive
test of local realism, but only if higher identification
efficiencies than in today experiments will be reached.



\vspace{8mm}

Since 1935, the celebrated paper of Einstein-Podolsky-Rosen \cite{EPR} suggested that Quantum Mechanics (QM) could be an incomplete theory, representing a statistical approximation of a complete deterministic theory in which observable values are fixed by some hidden variable.

In 1964 Bell \cite{Bell} discovered that any realistic Local Hidden Variable (LHV) theory must satisfy certain inequalities which can be violated in QM leading to the possibility of an experimental test of the validity of QM as compared to LHV \footnote{Bell theorem does not concern non-local hidden variables theories. For a recent progress toward an experimental verification of one of them see \cite{dbb} and Ref.s therein.}.

Since then, many experiments (mainly  based on entangled photon
pairs) have been addressed to test Bell inequalities
\CITE{asp,franson,type1,type2,nos}, leading to a substantial
agreement with standard quantum mechanics (SQM) and strongly
disfavouring LHV theories. However, so far, no experiment has yet
been able to exclude definitively such theories. In fact, up to
now one has always been forced to introduce a further additional
hypothesis \CITE{santos,garuccio}, due to the low total detection
efficiency, stating that the observed sample of particle pairs is
a faithful subsample of the whole. This problem is known as { \it
detection or efficiency loophole} and it has been shown that it
could be eliminated only by reaching a detection efficiency of $0.81$
when using maximally entangled states or $0.67$ for non-maximally
entangled ones \cite{eb}. Similarly the use of equalities
\cite{Hardy,GHZ} instead of Bell inequalities does not change the
situation \cite{garuccio}. Thus, searching for new experimental
configurations able to overcome the detection loophole is of
course of the greatest interest.

In the last years relevant progresses in this direction have been realised by using Parametric Down Conversion (a process where a UV photon decays in a pair of correlated photons of lower frequency inside a non-linear crystal) for generating entangled photon pairs with high angular correlation.
The generation of entangled states by PDC has replaced other techniques, such as the radiative decay of excited atomic states, as it was in the celebrated experiment of A. Aspect et al. \CITE{asp}, for it overcomes some former limitations and in particular allows a much larger efficiency. Many interesting experiments have been realised using such a technique \cite{franson,type1,type2,ou,nos,dem}. However, even if an efficiency of $30 \%$ \cite{type2} was approached, the limit of a detection loophole free test of LHVT remains far.

A recent experiment \cite{Win} based on the use of Be ions has reached very high efficiencies (around 98 \%), but in this case the two subsystems (the two ions) are not really separated systems during the measurement and the test cannot be considered a real implementation of a detection loophole free test of Bell inequalities, even if it represents a relevant progress in this sense.

Even if little doubts remain on the validity of the standard quantum mechanics, considering the fundamental importance of the question, the search for other experimental schemes for a definitive test of local realism is therefore of the utmost interest.

In the last years many papers \cite{BK} have been devoted to study
the possibility of realising such a test by the use of
pseudoscalar meson pairs as $K \bar{K}$ or $B \bar{B}$. If the
pair is produced by the decay of a particle at rest in the
laboratory frame (as the $\phi$ at $Da \phi ne$), the two
particles can be easily separated to a relatively large distance
allowing a space-like separation of the two subsystems and
permitting an easy elimination of the space-like loophole, i.e.
realising two completely space-like separated measurements on the
two subsystems (where the space-like separation must include the
setting of the experimental apparatuses too). A very low noise is
expected as well.

The idea is to use entangled states (i.e. non factorizable in
single particle states) of the form: \bea |\Psi \rangle = { | K^0
({\vec p})  \rangle | \bar K^0 ({- \vec p}) \rangle  - | \bar K^0
({\vec p}) \rangle | K^0({-\vec p})  \rangle \over \sqrt{2} } = &
\cr = { | K_L ({\vec p})  \rangle |  K_S ({-\vec p})  \rangle  - |
K_S ({\vec p})  \rangle | K_L ({-\vec p})  \rangle \over \sqrt{2}
} & \cr \label{psi} \eea where ${\vec p}$ is the momentum in the
centre of mass frame.

Claims that these experiments could allow the elimination of the detection loophole in view of the high efficiency of
particles detectors, have also been made. However, we have shown \cite{nosK} that due to the necessity of identifying the
state through some of its decays and since the decay channel can depend on the value of the hidden variables, the
detection loophole appears in this case as well.

In a recent paper \cite{Garb} a new scheme was proposed which seems to overcome the objections of Ref. \cite{nosK}.
Considering the possible large relevance of this result for future experiments,
 in this paper, after having discussed some general properties of K entangled states,
 we consider critically this proposal. Our result is that this scheme represents an interesting alternative
 test of LHVT, however it could  really overcome the detection loophole only if larger identification efficiencies of
 $K^0$ and $\bar K^0 $ than the ones available in nowadays experiments will be reached.

The main idea of Ref. \cite{Garb} is generating a non-maximally entangled state:

\be
|\Phi \rangle = { | K_S  \rangle | K_L  \rangle  - | K_L \rangle | K_S\rangle +
R | K_L  \rangle | K_L  \rangle  + R' | K_S \rangle | K_S\rangle
\over \sqrt{2+|R|^2+|R'|^2 } }
\label{psi2}
\ee
where $R=-r exp[-( i \Delta m + (\Gamma_S - \Gamma_ L)/2) T]$ and $R'=-r^2/R$, $\Delta m$ is the difference between $K_L$ and $K_S$ masses, whilst $\Gamma_S$, $\Gamma _L$ are their respective decay widths.

This can be obtained either by placing a regenerator slab on the path of kaons pairs produced in $\phi$ decays, where r is the regeneration parameter, or by considering kaons produced in $p \bar p$ annihilation at rest, where r measures the relative strength of p to s wave channels.

Selecting the $K_S$ surviving after $T = 10 \tau_S$ one obtains the non-maximally entangled state (neglecting the small $|R'| < 7 \cdot 10^{-3} |r|$)

\be
|\Phi \rangle = { R | K^0  \rangle | K^0  \rangle  + R | \bar K^0 \rangle | \bar K^0\rangle +
(2-R) | \bar K^0  \rangle |  K^0  \rangle  - (2+R) | K^0 \rangle | \bar K^0 \rangle
\over 2 \sqrt{2+|R|^2} }
\label{psi3}
\ee

Varying $R$ this state ranges from a maximally entangled one (for R=0) to various degree of non-maximal entanglement. For a pure state, entanglement can be quantified by looking to the von Neumann entropy, $S(\rho)= - Tr[\rho log_2 \rho]$, of the reduced density matrix $\rho$ of one of the two particles: a maximally entangled state corresponds to $S=1$ and a disentangled to $S=0$. The values of $S$ for the system of Eq. \ref{psi3} are plotted in Fig.1 in function of the real and imaginary parts of $R$ \footnote{For the state of Eq. \ref{psi2} at $T=0$ one has $S=0$ if $r$ could reach $\pm 1$.} . Incidentally, the possibility of realising different entangled states by manipulating the initial one is of large relevance for quantum information purposes (as studies on decoherence), since entanglement is the main resource for this field \cite{QI}.

Then one selects, with an appropriate choice of parameters, the case $R=-1$ (corresponding to $S=0.59$), for which QM predicts the probabilities of joint detection:

\bea P_{QM} (K^0 , \bar K^0) = \eta \eta'/12 &\cr P_{QM} (K^0 ,
K_L) = 0 &\cr P_{QM} (K_L , \bar K^0) = 0 &\cr P_{QM} (K_S ,  K_S)
= 0 &\cr \label{PQM} \eea
where $\eta$ and $\eta '$ are the
detection efficiencies for identifying $K^0$ and $\bar K^0$
respectively.

In a hidden variable model with distribution $\rho(a)$ of the
hidden variable (or variables set) $a$ the probability of
observing a $K^0$ to the left and $\bar K^0$ to the right is: \be
P_{LR}(K^0,\bar K^0) = \int da \rho(a) p_l(K^0|a) p_r(\bar K^0|a)
= \eta \eta'/12 \leq \int_{A_{0,\bar 0}} da \rho(a) \ee where $
p_l(K^0|a), p_r(\bar K^0|a)$ are the single kaon probabilities of
detecting a $K^0$ to the left and $\bar K^0$ to the right
respectively and $ A_{0,\bar 0}$ is the set of hidden variables
corresponding to a $K^0$ to the left and $\bar K^0$ to the right.

In a LHVT the necessity of reproducing Eq. \ref{PQM} requires that
if a $K^0$ ($\bar K^0$ ) is observed to the left (rigth) a $K_S$ propagates  to
the right (left). Thus one has $ p_l(K_S|a) =1, p_r( K_S|a)=1 $ if
$a $ belongs to $ A_{0,\bar 0}$. This is at variance with QM
predictions \ref{PQM}, since:

\be
P_{LR}(K_S,K_S) = \int da \rho(a) p_l(K_S|a) p_r( K_S|a)  \geq \int_{A_{0,\bar 0}} da \rho(a)
\label{PLR}
\ee

This result seems therefore to show that even if the detection efficiency of strangeness eigenstates is small, nevertheless LHVT can be tested without any additional hypothesis if the $K_S$, $K_L$ can be determined with perfect efficiency. Experimentally, this determination is realised by looking to the decay between time $T_0 =10 \tau_S$, where the state \ref{psi3} is produced, and time $T_1$ such that a negligible $K_L$ contribution to decays used for tagging $K_S$ is still expected in the interval.

Of course, this result would be of the largest relevance because of the possibility of realising this test at $\phi$ factories as $Da \phi ne$.

However, this discussion does not consider that in a deterministic theory hidden variables could also fix the channel of decay and the precise time of decay.

Therefore, Eq. \ref{PLR} becomes:
\be
P_{LR}(K_S, K_S) =\sum_C \int_{T_0}^{T_1} dt \sum_{C'} \int_{T_0}^{T_1} dt' \int da \, \rho(a) p_l(K_S|a) p_r( K_S|a) p_{l,C}(t|a) p_{r,C'}(t'|a)
\label{PLR2}
\ee
where $C$ and $C'$ run over the different decay channels (allowing an identification of $K_S$) and $ p_{i,C}(t|a)$ gives the probability of the $i=l,r$ (left, right) meson to decay into the channel $C$ at time t.

Let us now consider how this modifies the discussion concerning Eq.
\ref{PLR}.
 If the efficiencies $\eta$ and $\eta '$ of
$K^0$ and $\bar K^0$ detection were high, the situation would not
substantially change . However, unluckily, they are very small.
The method for this detection consists \cite{CLEO} in looking to
distinct interaction of $K^0$ and $\bar K^0$ with matter
(interaction and therefore identification that in principle could
depend on the hidden variables value): this led in Ref.
\cite{CLEO} to the identification of 70 unlike-strangeness events
and 19 like-strangeness events over $8 \cdot 10^7$ analyzed
events!
Thus, the few $K^0$-$\bar K^0$ identified events could easily
correspond to $K_S$ which would  not have decayed in the temporal
window that allows their identification and thus could not
contribute to the integral \ref{PLR2} (on the other hand if $\eta$
and $\eta '$ were sufficiently large this would not be possible). Furthermore,
it must also be considered that $| K_S \rangle$ and $| K_L \rangle
$ are not perfectly orthogonal, for $\langle K_S | K_L \rangle =
3.3 \cdot 10^{-3}$ \cite{PDB}. This means that a fraction of $K_S$
in the LHVT belongs to a hidden variable set corresponding to
decays characteristic of $K_L$ (and vice versa), giving  a further
fraction of states not contributing to $ P_{LR}(K_S,K_S) $ as
defined in Eq. \ref{PLR2}. Albeit very small this contribution
cannot be neglected due to the small value of $P_{LR}(K^0,\bar
K^0) = \eta \eta'/12$.

In summary, the small fraction of simultaneously identified left
$K^0$ and right $\bar K^0$ could be easily accounted for, in a
LHVT, by a fraction of $K_S$ that does not decay in an
identifiable form, since they decay outside of the temporal window
$T_0 < T < T_1$ or in an allowed channel for $K_L$, as three pions
or pion, lepton, neutrino.

From a numerical point of view, let us consider the optimistic case where
all kaons decayed before the time $T =10 \tau_s$ are identified
and therefore the surviving $K$ of a pair where one of members
decayed does not represent a background: the background
contribution from CP violating decays of $K_L$ to specific $K_S$
decays (used to tag them) becomes a $50 \%$ of the one due to {\it
true} $K_S$ contribution in the interval $21 \tau_S < t < 22
\tau_S$ and it is 1.35 times larger than $K_S$ contribution in the
interval $22 \tau_S < t < 23 \tau_S$. This means that $T_1$ must
be kept smaller than $\approx 21 \tau_S$.
 At $T=T_1= 21 \tau_S$ a fraction $1.5 \cdot 10^{-5} $ of the $K_S$ present at $T_0$ is still undecayed and a
 fraction $7.2 \cdot 10^{-4}$ is decayed in channels such that they could not be identified as $K_S$. This  gives the possibility of including in this set all the $K_S$ pairs that would correspond to identified left
 $K^0$ and right $\bar K^0$, if the identification efficiency of $K^0$, $\bar K^0$ is smaller of this amount, as for the experiment \cite{CLEO}.
Namely, since not all the $K_S$ are identified and since
hidden variables can be related also to decay channel and time (and therefore to the possibility of identifying the $K_S$), an experimental falsification of LHVT would require that
\be
P^{Measured} (K^0 , \bar K^0) = \eta \eta'/12 > 7.3 \cdot 10^{-4}.
\label{fals}
\ee
On the other hand, if inequality \ref{fals} is not verified, the hidden variable set corresponding to the observed $(K^0 , \bar K^0)$ could correspond to $K_S$ pairs which would not be observed as such and then do not contribute to an experimentally detectable difference between SQM and a LHVT.

The experimental realization of inequality \ref{fals} (i.e. $\eta
\approx \eta ' > 9 \%$) seems very difficult to be obtained.

A similar result can also be obtained, with a somehow different argument, rewriting the Equations
\ref{PQM} in the Clauser-Horne like inequality:
\be P (K^0 , \bar
K^0) > P (K^0 ,K_L) + P (K_L , \bar K^0) + P (K_S ,
K_S)
\label{CH}
\ee
and considering the possibility of a misidentification of a $K_S$ or $K_L$

The experimental measured probabilities become:
\bea P_{Measured} (K^0 , \bar K^0) = \eta \eta'/12
&\cr P_{ Measured } (K^0 , K_L) = (\eta m_S)/6 &\cr P_{ Measured } (K_L , \bar
K^0) = (\eta ' m_S)/6 &\cr P_{ Measured } (K_S ,  K_S) = {2 \over 3} m_L + {1 \over 3} m_L^2 &\cr \label{PQME}
\eea
where $m_S$ ($\approx 7.3 \cdot 10^{-4} $ from the previous discussion) and $m_L$ ($\approx 5.7 \cdot 10^{-5}$ due to two pions decays in the time interval $T_0 < t <T_1$) correspond to the probability of misidentification of
a $K_S$ or a $K_L$ respectively.

Inserting Eq. \ref{PQME} in \ref{CH} and assuming $\eta = \eta ' '$ it follows that an experimental violation of the inequality \ref{CH} requires $\eta = \eta' > 2.3 \cdot 10^{-2}$ (where the effect due to $m_L$ dominates).

 From the previous
discussion follows that  the observation of $ P_{LR}(K_S, K_S)=0$
in presence of a small $ P_{LR}(K^0,\bar K^0) \neq 0$ is not
sufficient for rejecting in general every conceivable LHVT if the
identification efficiency of $K^0$ and $\bar K^0$ is not
sufficiently high. On the other hand, in the proposed scheme, one
cannot hope to increase the $K^0$,$\bar K^0$ detection efficiency
by simply increasing strongly the width of the interposed
materials since the $K^0$,$\bar K^0$ measurement must be performed
in a small time interval around $T_0$ before the coefficient $R$
of the state \ref{psi3} changes substantially. Therefore, an
eventual experiment addressed to use this scheme for a conclusive
test of local realism should be very carefully planed in order to
satisfy all the discussed requests and in particular a high
$K^0$,$\bar K^0$ detection efficiency.

In conclusions we have shown that entangled neutral kaons states
represent a very interesting physical system for studying
entanglement and for stating severe limits to hidden variable
models. However, because of the results of Ref. \cite{nosK}, most
of the previous proposals \cite{BK} are not suitable for a
conclusive loophole free test of Local Hidden Variable Theories
against Standard Quantum Mechanics. Furthermore, the results of
this paper show that, considering the possible dependence of
channel and time of decay on hidden variables, the recent scheme
of Ref. \cite{Garb} could in principle led to a conclusive test,
but only if higher identification efficiencies of $K^0$ and $\bar
K^0$ will be reached.

\bigskip
\centerline{\bf Acknowledgements}
\bigskip
\noindent We would like to acknowledge support by  MURST contract
2001023718-002. We thank the anonymous referees for useful
suggestions.

\vfill \eject

\vskip 1cm

{\bf Figures Captions}

Fig.1   Three dimensional plot of the von Neumann entropy of the reduced density matrix in function of the real and imaginary parts of the parameter $R$ of Eq. \ref{psi3}.

Fig.2 Histogram of the ratio of the CP-violating contribution from
$K_L$ decays over the contribution of $K_S$ decays for pairs
surviving at $T=10 \tau_s$ and for bins of $1 \tau_s$ from $T=18
\tau_s$ to $T= 23 \tau_s$.
\end{document}